\documentclass[12pt]{iopart}
\usepackage{iopams}

\usepackage{amsfonts}
\usepackage{bm}

\usepackage{float}

\usepackage{hyperref}
\hypersetup{
	colorlinks=true,
	linkcolor=blue,
	filecolor=blue,
	citecolor=black,      
	urlcolor=cyan,
}

\usepackage{graphicx}
\graphicspath{{./}}

\usepackage{braket}

\usepackage{cite} 

\newcommand{\figref}[1]{figure~\ref{#1}}
\newcommand{\secref}[1]{section~\ref{#1}}
\newcommand{\eqnref}[1]{(\ref{#1})}

\newcommand{\expval}[1]{\left< #1 \right>}
\newcommand{\bigexpval}[1]{\big\langle #1 \big\rangle}

\newcommand{\deriv}[2]{\frac{\mathrm{d} #1}{\mathrm{d} #2}}

\newcommand{\mean}[1]{\overline{\langle#1\rangle}}

\newcommand{\erdosrenyi}{Erd{\H o}s-R\'enyi}
\newcommand{\ergraph}{G_{\mathrm{ER}}}

\newcommand{\hec}{h^\mathrm{e}_\mathrm{c}}
\newcommand{\hdcI}{h^\mathrm{d-I}_\mathrm{c}}
\newcommand{\hdcII}{h^\mathrm{d-II}_\mathrm{c}}
\newcommand{\Hmf}{H^{\mathrm{MF}}}
\newcommand{\Hmfp}{\Hmf_p}

\usepackage[T1]{fontenc}

\begin{document}

\title[DQPTs on random networks]{Dynamical quantum phase transitions on random networks}
\author{Tomohiro Hashizume\textsuperscript{1,*}, Felix Herbort\textsuperscript{1}, Joseph Tindall\textsuperscript{2},
   and Dieter Jaksch\textsuperscript{1,3}}
   \address{\textsuperscript{1}The Hamburg Centre for Ultrafast Imaging and Institute for Quantum Physics, 
   University of Hamburg, Luruper Chaussee 149, Hamburg 22761, Germany.}
   \address{\textsuperscript{2}Center for Computational Quantum Physics, Flatiron Institute, New York, New York 10010, USA.}
   \address{\textsuperscript{3}Clarendon Laboratory, University of Oxford, Oxford OX13PU, UK.}
   \address{\textsuperscript{*}Author to whom any correspondence should be addressed.}
   \ead{tomohiro.hashizume@uni-hamburg.de}
\vspace{10pt}
\begin{indented}
\item[]\today
\end{indented}

\begin{abstract}
   We investigate two types of dynamical quantum phase transitions (DQPTs) in the
   transverse-field Ising model on ensembles of \erdosrenyi{} networks of size $N$. 
   These networks consist of vertices connected randomly with probability $p$ ($0<p\leq 1$). 
   Using analytical derivations and numerical techniques, 
   we compare the characteristics of the transitions for $p<1$
   against the fully connected network ($p=1$). 
   We analytically show that the overlap between the wave function after a quench and the wave function of the fully connected network after the same quench deviates by at most $\mathcal{O}(N^{-1/2})$. 
   For a DQPT defined by an order parameter,
   the critical point remains unchanged for all $p$. 
   For a DQPT defined by the rate function of the Loschmidt echo, 
   we find that the rate function deviates from the $p=1$ limit near vanishing points of the overlap with the initial state, 
   while the critical point remains independent for all $p$.
   Our analysis suggests that this divergence arises from 
   persistent non-trivial global many-body correlations absent in the $p=1$ limit. 
\end{abstract}

\vspace{2pc}

\submitto{\NJP}

\maketitle

\section{Introduction}

\begin{figure*}[t]
    \centering
    \includegraphics[scale=1.0]{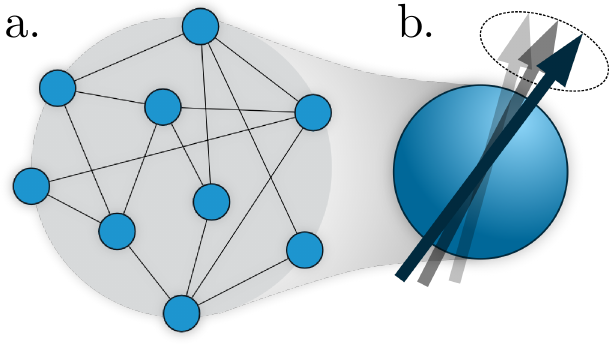}
    \caption{ 
       \textbf{Illustration of transverse-field Ising model on \erdosrenyi{} network.} 
       \textbf{a.~Instance of an \erdosrenyi{} network.}
       Blue dots represent individual spins at the vertices of a network, while black edges indicate interactions between spins in the Hamiltonian as given in equation \eqnref{eq:model_ising}.
       \textbf{b.~Convergence of local observables.}~Despite the missing links, 
       the wave function converges to the fully connected case ($p=1$) as $\mathcal{O}(N^{-1/2})$, 
       where the dynamics in this limit reduce to the oscillations of a large collective spin.
    }
    \label{fig:model}
\end{figure*}

Mean-field approximations have provided a powerful tool for understanding physical systems with high connectivity or large physical dimensions.
In many cases, these approximations prove to be exact in the thermodynamic limit, 
or at the very least, practically sufficient for capturing their essential physical properties 
\cite{Landau1969,pitaevskii1961vortex,grossStructureQuantizedVortex1961,
GLICK1965211,MESHKOV1965199,lipkinValidityManybodyApproximation1965,Amit1974,
negeleMeanfieldTheoryNuclear1982,botetLargesizeCriticalBehavior1983,staufferIntroductionPercolationTheory1994,
Cardy1996,kadanoffMoreSamePhase2009,sachdevQuantumPhaseTransitions2011,sendeEntanglementMeanField2012,
tindallQuantumPhysicsConnected2022,searleThermodynamicLimitSpin2024}.
However, a fundamental question remains: 
under what conditions does the mean-field approximation hold exactly, 
and when does it fail to capture the true dynamics of the system? 
Addressing this question is not only of theoretical interest but also of practical importance, 
as it deepens our understanding of the boundaries between simplicity and complexity in physical systems.

Recent technological advancements now allows us to engineer and probe large-scale artificial quantum systems 
with single-atom resolution
\cite{zhangObservationManybodyDynamical2017,brydgesProbingRenyiEntanglement2019,
   munizExploringDynamicalPhase2020,periwalProgrammableInteractionsEmergent2021,
weiObservationBraneParity2023,elbenRandomizedMeasurementToolbox2023,bluvsteinLogicalQuantumProcessor2024}.
These advancements enable experimental investigations of physical systems on artificial complex networks beyond those found in nature.
Examples of such systems include spin glasses 
\cite{sachdevGaplessSpinfluidGround1993,bovierThermodynamicsCurieWeissModel1993,
   panchenkoSherringtonKirkpatrickModelOverview2012,menonPhysicsDisorderedSystems2012,kabluchkoFluctuationsMagnetizationIsing2019},
chaotic systems \cite{hartmannChaosScramblingQuantum2019}, 
and toy models of quantum black holes
\cite{kitaevSimpleModelQuantum2015,bentsenTreelikeInteractionsFast2019,xuSparseModelQuantum2020,chenOperatorGrowthBounds2021},
where accessing their microscopic constituents is a difficult task. 

Furthermore, the recent advancement in computational technologies
has revealed the emergence of intelligence-like phenomena as a result of information propagation through complex networks
\cite{hopfieldNeuralNetworksPhysical1982,coleman2019selection,liuGPTUnderstandsToo2024,openaiGPT4TechnicalReport2024}.
These phenomena often deviate significantly from the dynamics predicted by mean-field theories, 
and the theoretical understanding of such emergent behaviour remains limited
\cite{okawaCompositionalAbilitiesEmerge2024,zhangIntelligenceEdgeChaos2024,chenQuantifyingSemanticEmergence2024}.
Developing theories to describe such systems is now an urgent priority, driving theoretical interest in recent years
\cite{andersonMoreDifferent1972,kaplanScalingLawsNeural2020,artimeOriginLifePandemics2022,
   barnettDynamicalIndependenceDiscovering2023,
greenEmergenceComplexNetworks2023}. 

Motivated by these recent developments, in this article, we study how strongly disordered interactions 
affect the dynamics of a quantum spin system.
We investigate this by analysing the properties of the dynamical quantum phase transitions (DQPTs)---the 
dynamical counterparts of equilibrium phase transitions
\cite{
   schutzholdSweepingSuperfluidMott2006,uhlmannVortexQuantumCreation2007,
   uhlmannSystemSizeScaling2010,
   itinDynamicsQuantumPhase2010,sciollaDynamicalTransitionsQuantum2011,  
heylDynamicalQuantumPhase2013,zunkovicDynamicalPhaseTransitions2016,  
homrighausenAnomalousDynamicalPhase2017,heylDynamicalQuantumPhase2018,  
zunkovicDynamicalQuantumPhase2018,lackiDynamicalQuantumPhase2019,  
heylDynamicalQuantumPhase2019,homrighausenOutEquilibriumMean2019,  
halimehQuasiparticleOriginDynamical2020,munizExploringDynamicalPhase2020,  
sunDynamicalQuantumPhase2020,halimehDynamicalPhaseTransitions2021,  
hashizumeDynamicalPhaseTransitions2022,marinoDynamicalPhaseTransitions2022}---in 
the transverse-field Ising model on \erdosrenyi{} networks $\ergraph(N,p)$ (TFIM-ER) 
\cite{erdds1959random,erdos1960evolution,bollobasCliquesRandomGraphs1976,bollobasModernGraphTheory1998,
newmanStructureFunctionComplex2003}. 
Those networks consist of $N$ vertices where each pair of vertices is connected with probability $p$ ($0<p<1$) 
as illustrated in \figref{fig:model}a. 

As our main result, we prove that in the thermodynamic limit, 
the time evolution of the overlap betwen the wave function in TFIM-ER and that of the $p=1$ (fully connected) limit converges to $1$.
This is shown by expanding the established duality between 
the equilibrium properties of TFIM-ER
and its $p=1$ counterpart \cite{tindallQuantumPhysicsConnected2022,searleThermodynamicLimitSpin2024}.
Consequently, the model's dynamics reduce to collective oscillations of local spins, as illustrated in \figref{fig:model}b,
and the dynamical critical points coincide with the analytically known critical points in the $p=1$ limit. 

However, our proof does not make statements on the $p$-dependence for the observables in finite size systems or 
for observables that are non-linear or non-local.
To address this, we perform finite size numerics using matrix product states (MPS)
\cite{perezgarciaMatrixProductState2006,mccullochDensitymatrixRenormalizationGroup2007,verstraeteMatrixProductStates2008,
   pirvuMatrixProductOperator2010,haegemanTimeDependentVariationalPrinciple2011,
   schollwockDensitymatrixRenormalizationGroup2011,
   orusPracticalIntroductionTensor2014,haegemanUnifyingTimeEvolution2016,
   paeckelTimeevolutionMethodsMatrixproduct2019,yangTimeDependentVariational2020,fishmanITensorSoftwareLibrary2022}, 
discretized semiclassical phase-space approach called a discrete truncated Wigner approximation (DTWA) 
\cite{schachenmayerManyBodyQuantumSpin2015,zhuGeneralizedPhaseSpace2019,czischekDiscreteTruncatedWigner2020},
and mean-field methods
\cite{polkovnikovPhaseSpaceRepresentation2010,langConcurrenceDynamicalPhase2018,homrighausenOutEquilibriumMean2019}. 
We confirm that the time evolutions of local observables converge to those of the fully connected model.
However, the rate function---a quantity that is analogous to free energy density in the thermal phase 
and highly non-linear in the wave function---deviates from the $p=1$ limit 
due to contributions from persistent global correlations in the $p<1$ system. 

The rest of this article is structured as follows:
In \secref{sec:ergraph}, we define the model, TFIM-ER. 
Then in \secref{sec:DQPT}, we briefly provide an overview of the theory of DQPTs 
and explain how different phases are characterized for the different types of the DQPTs. 
In \secref{sec:Results} we present our main result and the supporting numerical results obtained 
with MPS, DTWA, and mean-field simulations. 
Finally, we conclude and provide future outlook in \secref{sec:CandO}.

\section{Transverse-Field Ising Model on \erdosrenyi{} Network\label{sec:ergraph}}
The model we study in this article is the transverse-field Ising model on ensembles
of \erdosrenyi{} networks $\ergraph(N,p)$, which we refer to as TFIM-ER.
A network, $\ergraph(N,p) = \left(V_{p}, E_{p}\right)$, 
consists of $N$ vertices $V_p$ labelled $V_p\equiv\{ 0, 1, \cdots, N-1\}$, and $|E_p|$ edges. 
The variable $p$ ($0<p\leq 1$), 
dictates the probability of edge generation between every pair of vertices in the network.
The Hamiltonian of TFIM-ER on an instance of an \erdosrenyi{} network is given as
\begin{equation}
   H_p(h) = -\frac{J}{\mathcal{N}}\sum_{(i,j) \in E_p}  \sigma_i^z \sigma_j^z - h \sum_{i \in V_p} \sigma_i^x, 
    \label{eq:model_ising}
\end{equation}
where $\sigma^{x}$ and $\sigma^{z}$ are the dimensionless Pauli operators, and the Kac-normalization factor $\mathcal{N} = |E_p|/N$ 
is used to ensure that the energy-density is intensive
\cite{tindallQuantumPhysicsConnected2022,homrighausenAnomalousDynamicalPhase2017}.
We fix the Planck constant $\hbar=1$, and $J=1$ such that the interaction is ferromagnetic.
Furthermore, we restrict ourselves to $h>0$ as the results are symmetric about $h=0$ 
\cite{dasInfiniterangeIsingFerromagnet2006}. 

In the limit of $p=1$, we recover the fully connected network. 
Then, the model is exactly solvable, and known as the Ising limit of an anisotropic Lipkin-Meshkov-Glick (LMG) model
\cite{lipkinValidityManybodyApproximation1965,MESHKOV1965199,GLICK1965211,castanosClassicalQuantumPhase2006,
   ribeiroExactSpectrumLipkinMeshkovGlick2008,itinDynamicsQuantumPhase2010,sendeEntanglementMeanField2012}.
As a consequence, the Hamiltonian reduces to that of a non-interacting single large classical spin variable $\Theta$,
\begin{equation}
   H_{1} (h) = - \frac{J N^2}{(N-1)} \left(\braket{\Theta^{z}} \right)^2 - hN \braket{\Theta^{x}} + C, \label{eq:p1H}
\end{equation}
where $\Theta^{\alpha}=\sum_{i}^{N-1} \sigma^{\alpha}/N$ is the average spin operator. The constant term $C$ results from the self interaction and does not play a role in the dynamics.

The equilibrium phases of the Hamiltonian in \eqnref{eq:model_ising} for both $p<1$ and $p=1$ in the thermodynamic limit are well studied
\cite{castanosClassicalQuantumPhase2006,dasInfiniterangeIsingFerromagnet2006,
homrighausenAnomalousDynamicalPhase2017,aronLandauTheoryNonequilibrium2020,tindallQuantumPhysicsConnected2022}. 
They both have an equilibrium quantum critical point at $\hec=2$ for all $p$. 
The order of the ground state is characterized by the order parameter $\braket{\Theta^{z}}$
with associated $\mathbb{Z}_2$ symmetry. 
This symmetry breaks at the critical point. 
The phase diagram of the model is provided in \ref{sec:eqpd} for completeness. 

\section{Dynamical Quantum Phase Transitions}
\label{sec:DQPT}

Conventional equilibrium phase transitions are driven by control parameters such as an external field or temperature.
Analogous to the equilibrium case, DQPTs are induced by quenching a system parameter;
for TFIM-ER, this is an external field $h$.
Such a quench modifies the spectral structure of the Hamiltonian, which determines the dynamical phase of the system after the quench.
In this article, we investigate two approaches to defining DQPTs in TFIM-ER:
DQPT-I, based on the symmetry of the steady state, 
and DQPT-II, based on non-analyticities in the rate function of the Loschmidt echo, 
which serves as the dynamical analogue of free energy.

In quantum quench dynamics, the ground state $\ket{\psi_p(t=0)}$ 
of a Hamiltonian $H_p(h=h_i)$ is prepared for an initial transverse field strength $h_i$. 
For the quenches considered in this article, we fix $h_i=0$, 
and hence $\ket{\psi_p(t=0)}=\ket{\psi(0)}=\ket{\uparrow}^{\otimes N}$,
where $\ket{\uparrow}$ is the $+1$ eigenstate of $\sigma^{z}$. 
At $t=0$, the external field $h=h_i$ is changed abruptly to $h=h_f$.
Due to the change in the spectrum of the Hamiltonian, 
the state undergoes a time evolution.

Upon a quench, in DQPT-I, the critical field strength $h_f=\hdcI(p)$ marks a transition
between the symmetric and symmetry-broken phase in the time-averaged limit. 
For $h_f<\hdcI(p)$, a fraction of the initial order remains after time averaging. 
In contrast, above the critical point $h_f>\hdcI$, the initial order melts and does not survive.
In TFIM, DQPT-I is characterized by the relaxation of the $\mathbb{Z}_2$ symmetry 
with corresponding order parameter $\braket{\Theta^{z}}$. 

The phases in DQPT-II, on the other hand, 
are distinguished by the appearance of non-analytical cusps in the rate function
\begin{equation}
   \lambda(t) = -\frac{1}{N}\ln \left| \mathcal{G}(t) \right|^2,
   \label{eq:ratefunc}
\end{equation}
depending on the value of $h_f$. 
DQPT-II is motivated by the similarity between the canonical partition function $\mathcal{Z}$ 
(and thus the free energy density)
in statistical mechanics and the Loschmidt amplitude $\mathcal{G}$ in quantum mechanics
\begin{eqnarray}
   \mathcal{Z}(\beta) &= \Tr \{ e^{-\beta H_p(h)} \}\;,\\
   \mathcal{G}(t) &= \bra{\psi(0)} e^{-\mathrm{i}H_p(h)t}  \ket{\psi(0)}, \label{eq:loschmidt_amp}
\end{eqnarray}
where $\beta$ is the inverse temperature \cite{heylDynamicalQuantumPhase2013,heylDynamicalQuantumPhase2018}.

The phases in DQPT-II are, therefore, characterized by how non-analyticities appear in the dynamics of the rate function,
similarly to non-analyticities emerging in the free energy density $f(\beta)=-\frac{1}{\beta N} \ln \mathcal{Z}(\beta)$ 
at the critical temperature. 
In the regular phase ($h_f>\hdcII$), 
these non-analyticities periodically occur in the form of cusps. 
This phase often appears in the symmetric DQPT-I phase and 
the cusps are typically associated with the zero-crossings of the order parameter
\cite{heylDynamicalQuantumPhase2013,halimehDynamicalPhaseDiagram2017,hashizumeDynamicalPhaseTransitions2022}.

For $h_f<\hdcII$ in contrast, we expect contributions from the ordered initial state to survive, 
and therefore no cusps to appear.
This phase is referred to as the trivial phase and often associated with the symmetry-broken DQPT-I phase.
Nevertheless, in long-range models, the presence of non-analytical cusps are reported in the symmetry-broken DQPT-I 
phase due to the energetically favourable nature of local spin-flip excitations over domain-wall formations
\cite{halimehDynamicalPhaseDiagram2017,homrighausenAnomalousDynamicalPhase2017,
halimehQuasiparticleOriginDynamical2020,hashizumeDynamicalPhaseTransitions2022,vandammeAnatomyDynamicalQuantum2023}.
We refer to this phase as the anomalous phase. 
A key feature that is consistent across anomalous phases in different models is that the first cusp always appears 
after the first minimum of $\lambda(t)$.

In the $p=1$ limit, DQPT-I and DQPT-II in the TFIM-ER are well studied, and their nature is well understood
\cite{castanosClassicalQuantumPhase2006,dasInfiniterangeIsingFerromagnet2006,
homrighausenAnomalousDynamicalPhase2017,aronLandauTheoryNonequilibrium2020}. 
For the quenches considered in this article ($h=h_i=0~\rightarrow h=h_f$),
the dynamical critical points for both transitions lie at $\hdcI(p=1)=\hdcII(1)=1$.
The DQPT critical points can differ from equilibrium critical points
because DQPT critical points are determined by the full spectral properties of the Hamiltonian 
rather than those around the ground state. 
In this $p=1$ limit, $\hdcI$ and $\hdcII$ are related analytically to $\hec$ 
via the underlying classical phase space structure
that reflects the full spectral properties of the Hamiltonian in the thermodynamic limit
\cite{sciollaDynamicalTransitionsQuantum2011,homrighausenAnomalousDynamicalPhase2017}. 
Similar to the equilibrium counterpart, the DQPT-I critical point separates two phases given by the $\mathbb{Z}_2$ 
symmetry-breaking phase ($h<\hdcI(1)$) and the symmetric phase ($h>\hdcI(1)$). 
In contrast, the DQPT-II critical point separates the anomalous ($h<\hdcII(1)$) from the regular ($h>\hdcII(1)$) phase. 
In this limit, the dynamical critical points for DQPT-I and DQPT-II coincide.

In the following section, we present the results of DQPTs on TFIM-ER for $0<p<1$. 
Building on the $p$-independence of thermodynamic quantities in equilibrium 
\cite{tindallQuantumPhysicsConnected2022,searleThermodynamicLimitSpin2024},
we show that DQPT critical points persist at $\hdcI(p)=\hdcII(p)=\hdcI(1)=\hdcII(1)=1$,
leading to the same phases as in the $p=1$ case.
In the thermodynamic limit, the fluctuations induced by the underlying disordered lattice are suppressed,
resulting in the expected disappearance of $p$-dependence. 
Furthermore, we demonstrate that in the regular phase of DQPT-II, 
global correlations in the system survives, leading to 
qualitative distinctions in the behaviour of the rate function compared to the $p=1$ limit. 

\section{Results\label{sec:Results} and Discussions}

\begin{figure*}[t]
   \centering
   \includegraphics[scale=1.0]{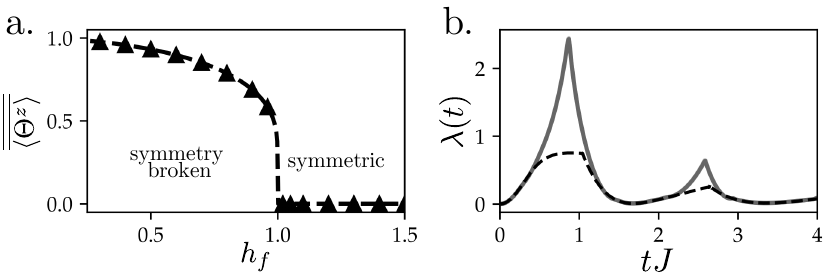}
   \caption{ 
      \textbf{Two ways of identifying dynamical quantum phase transitions (DQPTs).}
      \textbf{a.~DQPT-I}
      Plotted are the time-averaged $\mathbb{Z}_2$ order parameter $\overline{\mean{\Theta^z}}$
      of \erdosrenyi{} networks as a function of quench parameter $h_f$.  
      Triangles show the time-averaged value over the first 100 oscillations, 
      averaged over $100$ realizations of $\ergraph(5000,0.5)$ in the mean-field limit.
      The dashed line
      represents analytically computed values for
      the same quench for $p=1$ in the thermodynamic limit (\ref{app:timeaverage}). 
      Error bars are presented, although they are too small to be visible.
      \textbf{b.~DQPT-II}
      The rate function $\lambda(t)$ as a function of time $t$ 
      after a quench from $h_i=0$ to $h_f=2>\hdcII$ over 10 realizations of $\ergraph(100,0.5)$. 
      The dashed line represents the numerically exact (to 200 significant figures \cite{MATLAB}) function values for the same quench for $\ergraph(100,1)$. 
   }
   \label{fig:dqpts}
\end{figure*}

Our main result establishes a bound of $\mathcal{O}(N^{-1/2})$ on the divergence of fidelity between the time evolved states
$\ket{\psi_p(t)}=\exp(-iH_p(h_f) t)\ket{\psi(0)}$ and $\ket{\psi_1(t)}$
\begin{equation}
   \left| \deriv{}{t}\braket{\psi_p(t)|\psi_1(t)} \right| = 
   \Big|i\bra{\psi_p(t)}(H_p(h)-H_1(h))\ket{\psi_p(t)}\Big|
   = \mathcal{O}(N^{-1/2}),\label{eqn:derivbound}
\end{equation}
where $\ket{\psi(0)}=\ket{\uparrow}^{\otimes n}$ is the ground state of $H_p(0)$. 
The derivation of the above equation is provided in \ref{app:convdynamicsproof}.
This result proves that, in the thermodynamic limit, the parameter $p$ does not influence the behaviour of observables whose support is not extensive in the system size. 
This independence arises from the recovery of permutation symmetry over finite sets of vertices
in \erdosrenyi{} networks in the thermodynamic limit. 
In this limit, provided $p$ is finite and independent of the system size, the network converges almost surely to a Rado graph, 
a structure known for its permutation symmetries over any finite sets of its vertices 
\cite{erdosAsymmetricGraphs1963,cameronRandomGraphRevisited2001,tindallQuantumPhysicsConnected2022}. 

As a direct consequence of this bound, the critical point of DQPT-I in \erdosrenyi{} networks coincides with 
that of a fully connected network in the thermodynamic limit, as shown in \figref{fig:dqpts}a in the mean-field limit. 
In \secref{subsec:DQPT-I}, we validate this result using 
numerical simulations of $\braket{\Theta^{z}}$ with both fully quantum (MPS) and semiclassical (DTWA) methods. 
Our results show that the semiclassical approach accurately captures the quantum dynamics even for systems with as many as
$N=100$ spins.
Furthermore, the time-averaged value of the order parameter, 
\begin{equation}
   \overline{\mean{\Theta^{z}}}=\frac{1}{t_f}\int_0^{t_f} \mean{\Theta^z} dt, \label{eq:tavg}
\end{equation}
where $\mean{\cdots}$ denotes averaging over different network realizations,
converges to the value of the $p=1$ limit in the thermodynamic limit. 

Then, in \secref{subsec:DQPT-II}, 
we numerically analyse the dynamics of $\lambda(t)$ for various quench parameters $h_f$. 
Consistent with the bound, we obtain the same DQPT-II critical point that coincides
with the DQPT-I critical point ($\hdcII(p)=\hdcI(p)=1$). 
However, in the regular phase ($h_f>\hdcI(p)$), the dynamics of $\lambda(t)$ 
exhibit distinct behaviours for $p<1$ and $p=1$, 
especially near the turning points of the order parameter, where the overlap with the initial state vanishes
(\figref{fig:dqpts}b). 
In \secref{subsec:DQPT-II},
we further analyse the origin of these differences, 
identifying global correlations as a key factor influencing the rate function. 

\subsection{DQPT-I}
\label{subsec:DQPT-I}

\begin{figure}[t]
   \centering
   \includegraphics[width=1.0\textwidth]{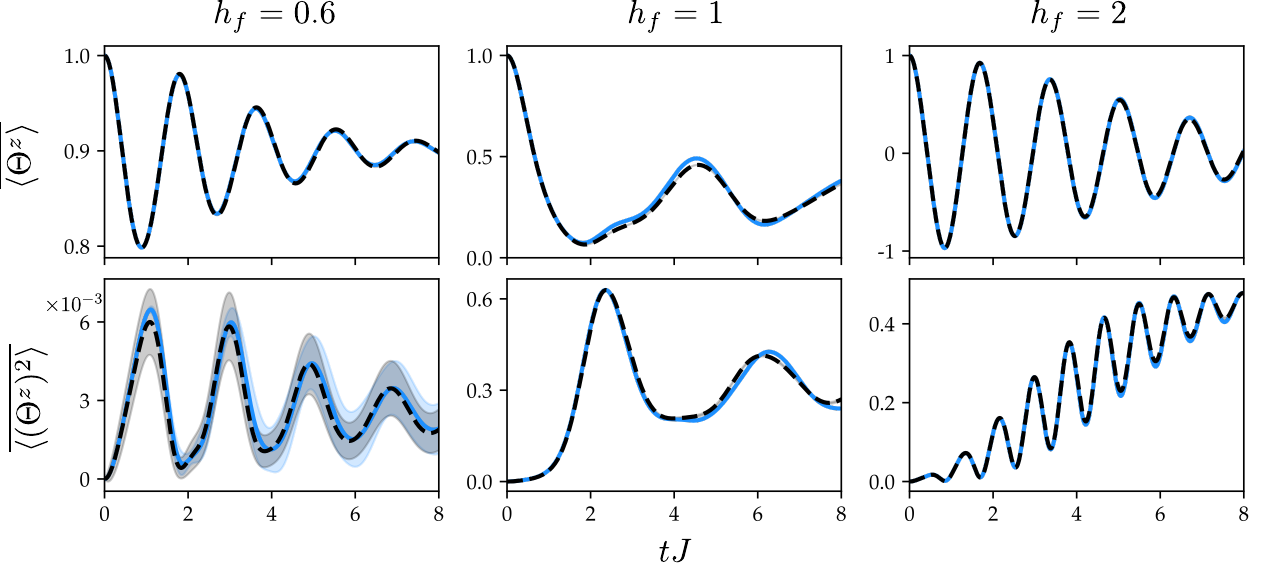}
   \caption{{ \bf Time evolution of the order parameter and its variance.} 
      The order parameter $\mean{\Theta^{z}}$ (top)
      and its variance $\mean{(\Theta^z)^2}$ (bottom) 
      are plotted for the quenches from $h_i=0$ to $h_f=0.6$ (left), $h_f=1.0$ (middle), and $h_f=2.0$ (right)
      for $p=0.5$. 
      Fully quantum results (blue solid) 
      are computed using the TDVP algorithm \cite{yangTimeDependentVariational2020,fishmanITensorSoftwareLibrary2022}
      with the bond dimension $\chi=200$ and time step $\Delta t=0.01$.
      The quantities are then averaged over 100 network realizations. 
      These are compared to the equivalent quenches that are simulated with semiclassical DTWA algorithms (black dashed),
      where the quantities are averaged over 100 trajectories per network for 100 network realizations. 
      At the dynamical critical point ($h_f=\hdcI=1$), small deviations are observed
      due to instabilities near the dynamical critical point (middle panels).
      Error bars are shown as shaded regions around the lines. Apart from the bottom left plot, 
      they are too small to be visible. 
   \label{fig:tdvpdtwa_comparison}}
\end{figure}

We first present the time evolution of the order parameter $\mean{\Theta^z}$ 
in the fully quantum limit for $N=100$, computed using the 
time dependent variational principle (TDVP) algorithm using matrix product states (MPS)
\cite{yangTimeDependentVariational2020,fishmanITensorSoftwareLibrary2022}.
An exact matrix product operator representation of the Hamiltonian is 
constructed from the linear array of long-range interactions,
following the approach used in \cite{tindallQuantumPhysicsConnected2022} for simulating equilibrium phases.
The maximum bond dimensions are kept to at most 102 for all simulations presented in this work.
These results are compared to those obtained from the discrete truncated Wigner approximation (DTWA), 
a semiclassical Monte-Carlo simulation on discretized phase space
\cite{schachenmayerManyBodyQuantumSpin2015,zhuGeneralizedPhaseSpace2019}.
As shown in the left and right panels of \figref{fig:tdvpdtwa_comparison}, 
the DTWA closely matches MPS for both the order parameter $\mean{\Theta^z}$
and its variance $\mean{(\Theta^z)^2}$ for all values of $h_f$. 
Importantly, DTWA avoids the pathological quadratic divergence over time that is observed in truncated Wigner approximation
calculations on continuous phase space for the $p=1$ limit 
\cite{homrighausenOutEquilibriumMean2019}.
At the DQPT-I critical point, 
a small deviation between DTWA and MPS results is observed (middle panel)
but it remains well-controlled within the simulation timescales. 

Based on the excellent agreement between the MPS and DTWA simulations,
we extend our analysis to larger network sizes and times beyond MPS capabilities.
We treat the DTWA trajectories of $\braket{\Theta^{z}}$ 
as an accurate approximation of the exact quantum dynamics.
We then analyse the underlying phase-space structure of the model, comparing the semi-classical and the classical limit
to gain further insight into the system's dynamics. 

\begin{figure}[t!]
   \centering
   \includegraphics[scale=1.0]{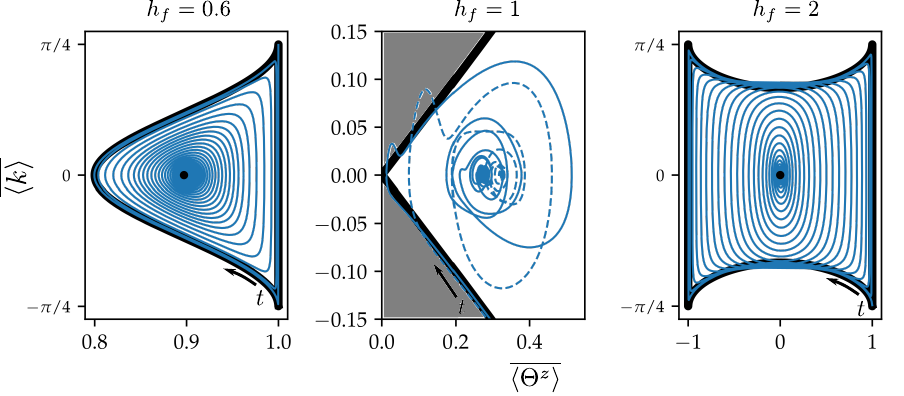}
   \caption{ \label{fig:meanfield}
      {\bf Phase space trajectories for an initially polarized state in the semiclassical regime.}
      Phase space trajectories of quenches, calculated using DTWA by averaging 10,000 trajectories, are shown in blue
      on \erdosrenyi{} networks with $p=0.5$ and $N=1000$ for $h_f=0.6$ (left), $1$ (middle), and $2$ (right)
      up to $t=100$ for 100 network realizations.
      For $h_f=1$, the trajectory for $N=100$ is plotted in dotted lines for comparison,
      and the classically forbidden region is indicated by shaded gray areas.
      The black thick lines represent the mean-field trajectories in the thermodynamic limit, 
      which are equivalent to the trajectories of TFIM-ER in the $p=1$ limit.
   }
\end{figure}

First, we obtain the effective classical mean-field Hamiltonian 
\begin{eqnarray}
   \Hmfp(h) =  
   - \frac{JN}{|E_p|} \sum_{i,j \in E_p } \braket{\sigma^z_i}\braket{\sigma^z_j} 
   - h \sum_{i} \sqrt{1-\braket{\sigma_i^z}^2}\cos 2 k_i,
\end{eqnarray}
for phase space variables $\braket{\sigma^z_i}$ 
and their conjugate momenta $k_i$ ($i\in\{0,1,\cdots,N-1\}$),
with $\braket{\sigma^{x}_i}=\sqrt{1-\braket{\sigma^{z}_i}^2}\cos 2 k_i$ 
and $\braket{\sigma^{y}_i}= -\sqrt{1-\braket{\sigma^{z}_i}^2}\sin 2 k_i$ (\ref{app:mfem}).
In \figref{fig:meanfield}, we show the phase space trajectories, 
parameterized by the averages of the phase space variables, $\mean{\Theta^{z}}$ and $\mean{k}=\mean{\sum k_i/N}$, 
obtained numerically with DTWA for $N=1000$. 
In the thermodynamic and $p=1$ limit of the model (black lines), there exist two distinct phase-space regions separated 
by a separatrix, corresponding to the trajectory for $h_f=\hdcI(1)=1$ (the black line in the middle panel).
This separation corresponds to a trajectory in the $p=1$ limit that exhibits a diverging orbital period.
The trajectory passes through an unstable fixed point at $(\braket{\Theta^z},k)=(0,0)$,
which separates $\mathbb{Z}_2$-invariant states from the rest
and is the origin of the DQPT-I in the model in the $p=1$ limit
\cite{sciollaDynamicalTransitionsQuantum2011,homrighausenAnomalousDynamicalPhase2017}. 

As shown in \figref{fig:meanfield},
for finite $N$, the trajectories of $\braket{\Theta^z}$ 
deviate from the $p=1$ limit due to fluctuations in $\braket{\sigma^{z}_i}$ and $k_i$. 
These fluctuations cause $\braket{\Theta^z}$ to exhibit damped oscillations around the time-averaged 
value of the of the $p=1$ limit
\begin{equation}
   (\overline{\mean{k}}_{p=1},\overline{\mean{\Theta^{z}}}_{p=1})=
   \left\{
      \begin{array}{ll}
      \left(0,\frac{\pi}{2K((h_f/J)^2)} \right) &(h_f < 1) \\
      (0,0)  &(h_f > 1)
      \end{array}
      \right.,
\end{equation}
where $K(m)$ is the elliptic integral of the first kind (cf.~\ref{app:timeaverage} for derivation).
Near the critical field strength $\hdcI=1$, these deviations are most pronounced, as shown in \figref{fig:meanfield} (middle).
Here, strong finite-size effects cause the steady-state values of the phase space variables 
to deviate from their thermodynamic limit values. 
Additionally, the trajectories enter a classically forbidden region, indicated in grey. 
However, in the thermodynamic limit, the trajectories of the $p=1$ limit become exact, 
as the wave functions converge to the $p=1$ limit, recovering the phase diagram in \figref{fig:dqpts}a. 

In summary, the critical point in the thermodynamic limit is determined by the properties of the $p=1$ limit of TFIM-ER.
As derived in equation \eqnref{eqn:derivbound}, the critical point occurs at $\hdcI(p)=\hdcI(1)=1$, independent of $p$, 
as shown in \figref{fig:dqpts}a.
This critical point divides the symmetry-breaking phase ($h_f<\hdcI(p)$) from the symmetric phase ($h_f>\hdcI(p)$).
This transition arises from the phase space structure of the $p=1$ limit, which dictates the system's critical behaviour. 

\subsection{DQPT-II}
\label{subsec:DQPT-II}

For $p=1$, previous studies have showed that
the model transitions from an anomalous phase ($h_f<\hdcII(1)=1$) to a regular phase $h_f>\hdcII(1)=1$
\cite{homrighausenAnomalousDynamicalPhase2017,homrighausenOutEquilibriumMean2019,halimehQuasiparticleOriginDynamical2020}.
However, 
It is not clear 
whether TFIM-ER possesses the same phases and DQPT-II critical point in the thermodynamic limit
due to the strongly non-local and nonlinear nature of $\lambda(t)$. 
To address this, we numerically investigate DQPT-II by computing the disordered averaged rate function 
\cite{yinZerosLoschmidtEcho2018,beniniLoschmidtEchoSingularities2021,vanhalaTheoryLoschmidtEcho2023}
\begin{equation}
   \overline{\lambda(t)} = -\frac{1}{N} 
   \ln \mean{\left| \mathcal{G}(t) \right|^2}
   \label{eq:ratefunc_avg}
\end{equation}
where $\mathcal{G}(t)$ is the Loschmidt amplitude as given in equation \eqnref{eq:loschmidt_amp}.

In this subsection, we show the critical point and phases for DQPT-II align with those in the $p=1$ limit. 
However, global correlations lead to qualitatively different behaviour of the rate function in the regular phase. 

\begin{figure}[t!]
   \centering
   \includegraphics[width=1.0\textwidth]{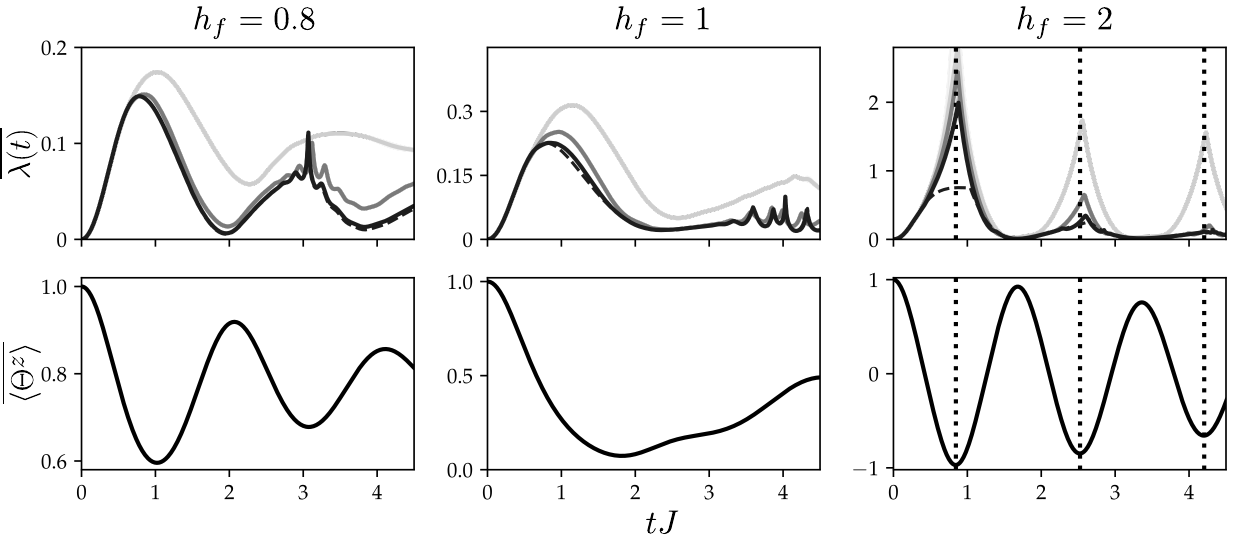}
   \caption{
      {\bf The rate function of Loschmidt echo.} 
      The rate function $\overline{\lambda(t)}$ 
      calculated from the average overlap of 100 trajectories with fixed $N=100$, 
      for $p=0.1$, $0.5$, and $0.9$ (light to dark) 
      for $h_f=0.8$ (left), $1$ (middle), and $2$ (right). 
      The dashed lines are $\overline{\lambda(t)}$ for $p=1$ and $N=100$ (exact). 
      Plotted in the bottom panels are $\mean{\Theta^{z}}$ for $p=0.5$; 
      for $h_f=2$, the turning points of $\mean{\Theta^{z}}$ are indicated with vertical dotted lines. 
      The simulations are conducted with TDVP algorithm with MPS with the maximum bond dimension $\chi=200$ and $\Delta t=0.01$. 
      The rate function is computed from the numerically obtained Loschmidt amplitude $\mathcal{G}(t)$.
      The error bars are plotted as shaded regions, but they are too small to be visible. 
      For $p=1$, $\lambda(t)$ is calculated exactly to 200 significant figures \cite{MATLAB}.
   \label{fig:LErr_mps}}
\end{figure}

Figure \ref{fig:LErr_mps} displays $\overline{\lambda(t)}$
for quenches to the ordered phase ($h_f=0.8$), critical regime ($h_f=1$), and the disordered phase ($h_f=2$) 
The results are shown for $p=0.1$, $0.5$, and $0.9$ (solid lines, light to dark), 
alongside the rate function for the fully connected network (dashed). 
For $p=0.5$ and $p=0.9$, 
cusps are observed near $t=3.1$ in the ordered phase (left panel), 
consistent with the $p=1$ limit (dashed line).
At the critical point ($h_f=1$), the rate function shows the cusps within a plateau
following the first maximum (middle panel). 
Additionally, as $N$ increases, the rate function converges towards the $p=1$ limit. 
(cf.~\ref{app:lambda_vs_N}). 

While cusps are observed for $p=0.5$ and $p=0.9$, they are absent for $p=0.1$ in quenches to the ordered phase. 
We identify the origin of the observed discrepancy for $p=0.1$ to strong finite-size effects, 
arising from the underlying structure of the \erdosrenyi{} network. 
When $p$ is sufficiently small, of order $\ln N /N$, 
the network possesses a chain-like structure with only a few small loops 
\cite{friezeIntroductionRandomGraphs2015,hofstadRandomGraphsComplex2016}.
For $N=100$ the threshold occurs at $\ln N /N\approx 0.05$.
Since $p=0.1$ is close to this threshold, the underlying geometry is dominated strongly by 
the tree-like geometry with local short-range interactions,
where the transverse-field Ising model is known to exhibit a trivial phase 
below the DQPT-II critical point \cite{halimehQuasiparticleOriginDynamical2020}.
For larger network sizes, we expect the anomalous cusps to reappear. 

Finally, the rightmost panel of \figref{fig:LErr_mps} shows that quenches deep into the disordered phase
produce rate functions with periodic cusps, characteristic of the regular phase. 
Unlike the $p=1$ limit, the cusp formation times
align with the lower turning points of the order parameter 
$\mean{\Theta^z}$ 
with the cusps becoming significantly sharper with increasing system size (\ref{app:lambda_vs_N}). 
This deviation suggests that, while the phase is the same for $p<1$ and $p=1$, it emerges from fundamentally different
mechanisms in the two cases. 
In the next section, we provide a detailed analysis of the regular phase to further explore the origin of 
the deviation from the $p=1$ limit and the alignment of the point of divergence and the turning point of the order parameter,
focusing on the role of global correlations in the system's dynamics. 

\subsection{The Regular Phase in DQPT-II}

Lastly, we demonstrate that the observed periodical divergence of the rate function in the regular phase 
is a distinctive feature of TFIM-ER ($p<1$). 
This behaviour stems from non-trivial global many-body correlations that survive in the thermodynamic limit.
We first define
\begin{eqnarray}
   \overline{C}_m
            = \frac{\mean{\Delta^{zzz\cdots}_{i,j,k\cdots} 
      \prod_{l\neq \{ i,j,k\cdots\} }\braket{ \frac{\sigma^z_l + \mathbb{I}}{2}}}
      }{\mean{\prod_{q}  \braket{ \frac{\sigma^z_q + \mathbb{I}}{2}}}}
\end{eqnarray}
where 
$\Delta^{zzz\cdots}_{i,j,k\cdots} = \braket{(\sigma^{z}_i - \braket{\sigma^{z}_i})(\sigma^{z}_j 
- \braket{\sigma^{z}_j})(\sigma^{z}_k - \braket{\sigma^{z}_k})\cdots}$ 
is the $m$\textsuperscript{th} order joint central moment over $m$ non-overlapping sites $i,j,k,\cdots$. 
Then, $\mean{ |\mathcal{G}(t)|^2 }$ is written as a sum of $\overline{C}_m$ as follows
\begin{eqnarray}
   \mean{|\mathcal{G}(t)|^2}  = 
   \mean{ \prod_{i}  \braket{ \frac{\sigma^z_i + \mathbb{I}}{2}}}
         \left(
      1 + \sum_m \frac{1}{2^{m}}\sum \overline{C}_m
         \right),
   \label{eqn:rfapprox}
\end{eqnarray}
where the second summation goes over all possible combinations of $m$ non-overlapping sites. 

When the higher-order moments $\Delta^{zzz\cdots}_{i,j,k\cdots}$ and 
their product $\prod_{l\neq \{ i,j,k\cdots\} }\braket{ \frac{\sigma^z_l + \mathbb{I}}{2}}$ are uncorrelated, 
$\mean{ |\mathcal{G}(t)|^2 }$, and hence $\overline{\lambda(t)}$, 
converges to the value of the $p=1$ limit in the thermodynamic limit.
However, for $p<1$, correlations emerge between these terms due to the underlying disorder in the network. 
Assuming that the distributions are approximately normal and log-normal, respectively, 
the mean of the product shifts by $\sigma_{0,m}$ from the $p=1$ limit, 
where $\sigma_{0,m}$ is the covariance between the distributions \cite{yangNormalLognormalMixture2008}.
The validity of this assumption is confirmed for the quenches analysed in this section and presented in \ref{app:flucandcumul}.

\begin{figure}
   \centering
   \includegraphics[scale=1.0]{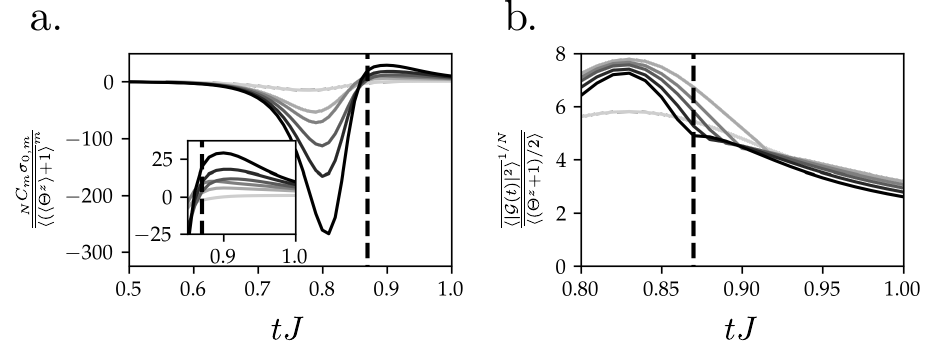}
   \caption{
      {\bf Contribution of the underlying disorder to the Loschmidt echo.}
      {\bf a.}~Evolution of the 
      contribution of the global correlations between $\Delta^{zz}_{i,j}$ and $\prod_{k\neq\{i,j\}} \braket{(\sigma^{z}_k+1)/2}$, 
      $\Delta_m \approx \frac{~_NC_m \sigma_{0,m}}{\mean{(\braket{\Theta^z}+1)}^m}$ for $m=2$,
      estimated from the statistical distribution of the two quantities. 
      Here $\sigma_{0,2}$ is computed after resolving it by the presence and absence 
      of the edge between vertices $i$ and $j$ (cf.~\ref{app:flucandcumul} for details).
      {\bf b.}~Contribution of the global correlations to the Loschmidt amplitude quantified by 
      $\frac{\mean{|\mathcal{G}(t)|}^{1/N}}{\mean{(\Theta^z+1)/2}}$, after removing contributions from a
      trivial product of local expectation values 
      $(\prod_i \braket{(\sigma^{z}_i+1)/2})^{1/N} \approx \braket{(\Theta^{z}+1)/2}$. 
      Dashed vertical lines in both panels are the lower turning points of $\mean{\Theta}$ for $N=100$.
      Both results are obtained by simulating the dynamics over
      100 network realizations for $p=0.5$ for quenches with $h_f=2$. 
      Different shades of black correspond to different system sizes ($N=20$, $40$, $50$, $75$, and $100$, light to dark).
   \label{fig:divergingcorrelation}}
\end{figure}

To further explore the implications of $\overline{C}_m$, 
we analyse the shift 
\begin{eqnarray}
   \Delta_m &= \frac{1}{2^m}\sum \overline{C}_m - \frac{1}{2^m}\sum 
      \frac{\mean{\Delta^{zzz\cdots}_{i,j,k\cdots}}
      \mean{\prod_{l\neq \{ i,j,k\cdots\} }\braket{ \frac{\sigma^z_l + \mathbb{I}}{2}}}
   }{\mean{\prod_{q}  \braket{ \frac{\sigma^z_q + \mathbb{I}}{2}}}} \nonumber\\
   & \approx \frac{~_NC_m \sigma_{0,m}}{\mean{(\braket{\Theta^z}+1)}^m}
   \label{eqn:Deltam}
\end{eqnarray}
of the overall sum $\sum \overline{C}_m$ from the $p=1$ limit, for $m=2$ for the quench with $h_f=2$ and $p=0.5$.
The summations go over all non-overlapping combinations of the sites 
that contribute to $\Delta^{zzz\cdots}_{i,j,k\cdots}$ and $~_NC_m=\frac{N!}{m!(N-m)!}$ is the binomial coefficient.
This is plotted in \figref{fig:divergingcorrelation}a. 
Notably, even for $m=2$, the contribution from the global many-body correlations survives towards the thermodynamic limit.
Especially, near the time of the first cusp (vertical dashed line), two diverging behaviours emerge:
one arises from the denominator of $\overline{C}_m$ approaching $0$ as $\mathcal{O}(N^{-m})$
as $\mean{\Theta^{z}}$ approaches $-1$, and the other arises from the abrupt change in the sign of $\sigma_{0,m=2}$ (inset).
This phenomenon is further evidenced by the emergence of a nonanalyticity and 
   the non-divergent nature of 
   $\frac{\mean{|\mathcal{G}(t)|}^{1/N}}{\mean{(\Theta^z+1)/2}}\approx\left(1+\sum_{m=2}\frac{1}{2^m}\sum \overline{C}_{m}\right)^{1/N}$ 
plotted in \figref{fig:divergingcorrelation}b. 
The observed system size dependence indicate that
   $\lim_{N\to\infty} \mean{|\mathcal{G}(t)|^2}^{1/N}=0$ as 
   $\lim_{N\to\infty}\mean{(\Theta^z+1)/2}=0$, hence the aligned divergences of $\overline{\lambda(t)}$ 
   with the lower turning points of $\mean{\Theta^z}$ (\ref{app:flucandcumul}). 

Building on the discussion of global correlations, we now examine their implications for the rate function in different phases.
For quenches where $\braket{\Theta^z+1}$ does not vanish ($h_f<\hdcI$), 
the contribution from $\sigma^{0}\sigma^{0,m}$ vanishes in the thermodynamic limit due to rapidly vanishing fluctuations. 
Thus, the rate function converges to the $p=1$ limit. 
For quenches into the symmetric phase ($h<\hdcI$), the scaling $\mean{\Theta^{z}+1}^m = \mathcal{O}(N^{-m})$ 
near $\mean{\Theta^{z}}=-1$ amplifies the global correlations, leading to sharp divergence of $\lambda(t)$ 
near the lower turning points of $\braket{\Theta^{z}}$, where the overlap with the initial state vanishes. 
Consequently, the critical points of both DQPT-I and DQPT-II lie at $\hdcI(p)=\hdcII(p)=1$ for the described quenches.

\section{Conclusions and Outlook\label{sec:CandO}}
In this work, we studied dynamical quantum phase transitions in the quench dynamics of the transverse-field Ising model 
on an ensemble of \erdosrenyi{} networks.
Building on the equilibrium case \cite{tindallQuantumPhysicsConnected2022},
we have proven analytically that the time derivative of $\braket{\psi_p(t)|\psi_1(t)}$ is bounded by $\mathcal{O}(N^{-1/2})$. 
Through numerical simulations, 
we further established that the dynamical critical points for both DQPT-I and DQPT-II are independent of $p$,
with transitions occurring at $\hdcI(p)=\hdcII(p)=1$.

While the dynamical phases of the model exhibit duality with its $p=1$ limit, 
a notable qualitative difference arises in the regular phase of DQPT-II. 
Specifically, the model shows strong divergence near the turning points of $\braket{\Theta^z}$, 
where the overlap with the initial state vanishes. 
We attribute this deviation from the $p=1$ limit to the influence of global correlations within the system. 
However, our analysis has so far focused only on the lowest-order contributions.
Future work should explore higher-order fluctuations and their implications for macroscopic phenomena. 

Future work could extend to random networks, such as small-world networks
\cite{albertStatisticalMechanicsComplex2002,newmanStructureFunctionComplex2003,
friezeIntroductionRandomGraphs2015,hofstadRandomGraphsComplex2016}
and \erdosrenyi{} networks with a system size dependent $p$,
which exhibit different automorphism characteristics in the thermodynamic limit than Rado or complete graphs. 
Such investigations could pave the way for developing a theory of defect detection in non-trivial quantum networks, 
including complex quantum circuits.
Additionally, the dynamics explored in this study could be experimentally probed using 
near-term quantum simulation platforms, 
such as cavity quantum electrodynamcis 
\cite{periwalProgrammableInteractionsEmergent2021,vaidyaTunableRangePhotonMediatedAtomic2018}, 
trapped ions 
\cite{Bohnet2016,richermeNonlocalPropagationCorrelations2014,mosesRaceTrackTrappedIonQuantum2023}, 
and atoms in optical lattices 
\cite{suDipolarQuantumSolids2023}. 
These platforms are known to support long-range interactions 
(cf.~\cite{defenuLongrangeInteractingQuantum2023} for a comprehensive review), though implementing randomly placed (cut-out) 
long-range interactions in a scalable manner remains a technological and experimental challenge.
Meanwhile, for moderately-sized \erdosrenyi{} networks, 
current trapped-ion devices (e.g.~\cite{mosesRaceTrackTrappedIonQuantum2023}) 
could be used to address the discrete-time dynamics of the Ising model, and shed light on how the lack of 
an underlying Hamiltonian generator of the dynamics affects the mean-field nature of the model.

\section{Acknowledgements}
JT is grateful for ongoing support through the Flatiron Institute, a division of the Simons Foundation.
DJ acknowledges support by the European Union's Horizon Programme (HORIZON- CL42021DIGITALEMERGING-02-10) Grant Agreement 101080085 QCFD, DFG project 'Quantencomputing mit neutralen Atomen' (JA 1793/1-1, Japan-JST-DFG-ASPIRE 2024), and the Hamburg Quantum Computing Initiative (HQIC) project EFRE.
The EFRE project is co-financed by ERDF of the European Union and by 'Fonds of the Hamburg Ministry of Science, Research, Equalities and Districts (BWFGB)'.
Authors from University of Hamburg are partly funded by the Cluster of Excellence 'Advanced Imaging of Matter' of the Deutsche Forschungsgemeinschaft (DFG)|EXC 2056- project ID390715994.
Results were obtained using the PHYSnet computational cluster based at University of Hamburg.
TH and FH acknowledge Martin Stieben for support on obtaining the numerical results.
Matrix product state calculations were performed with the \texttt{C++}-based software library
\texttt{ITensor} and its \texttt{TDVP} package \cite{fishmanITensorSoftwareLibrary2022,yangTimeDependentVariational2020}.
The data that support the findings of this study are available upon reasonable request from the authors and at reference \cite{hashizumedataset2025}.

\section*{References}
\bibliographystyle{iopart-num}
\bibliography{./bibliography.bib}

\newpage
\appendix
\section{Equilibrium Phase Diagram \label{sec:eqpd}}
\begin{figure}[t!]
   \centering
   \includegraphics[scale=0.8]{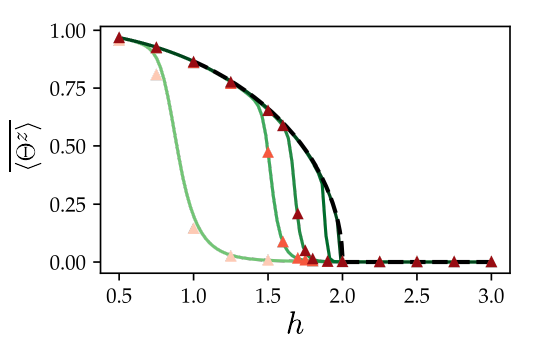}
   \caption{ {\bf Equilibrium phase diagram of \erdosrenyi{} network.} 
      The average ground state order ($\mean{\Theta^{z}}$) of the model 
      for $p=0.5$ and $N=10,50,100$
      (red triangles, light to dark). 
      They are computed with 100 realizations of \erdosrenyi{} network with the DMRG algorithm with maximum bond dimension $\chi=200$. 
      The results shows excellent agreement with their fully connected counterparts that are computed exactly
      ($p=1$, green lines for $N=10$, $50$, $100$, $500$, and $5000$ light to dark). 
      However, near the critical point $\hec{}(p=1)=2$, they both show disagreement with the analytically tractable values 
      of the order parameter in the thermodynamic limit 
      (equation \eqnref{eq:TLop}, black dashed line).
   }
   \label{fig:eqPD}
\end{figure}

In the limit of $p=1$, the model has a quantum equilibrium phase transition critical point 
at $\hec(p=1)=2$ \cite{dasInfiniterangeIsingFerromagnet2006}, where the ground state phase of the model transitions from 
ferromagnetic ($h<\hec{}$) to disordered phase ($\hec<h$). 
We derive this by minimizing the classical energy given by equation \eqnref{eq:p1H} 
with respect to the continuous phase space variable 
${\bm \Theta}=(\Theta^{x}, \Theta^{y}, \Theta^{z})$
where $\Theta^{\kappa} \in [-1,1]$ and $\sum_{\kappa} \big(\Theta^{k}\big)^2= 1$.
The phase transition is associated with the spontaneous $Z_2$ symmetry breaking 
with an order parameter $\braket{\Theta^{z}}$. 
This order parameter behaves like 
\begin{equation}
   \braket{\Theta^{z}} = 
   \left\{   
      \begin{array}{@{}l@{\quad}l}
         \sin \left( \arccos \left(h/2\right) \right) &(h\leq 2) \\ [\jot]
         0 &(h>2) 
      \end{array}
   \right. \label{eq:TLop}
\end{equation}
as a function of $h$.
Equation \eqnref{eq:TLop} is plotted in \figref{fig:eqPD} as a black dashed line. 

For $p<1$, we use the results from \cite{tindallQuantumPhysicsConnected2022,searleThermodynamicLimitSpin2024}
that showed the equivalence between the equilibrium critical point of \erdosrenyi{} network
for $p<1$ and $p=1$ (fully connected network).
This result comes from the convergence of the two-body interaction energy of any normalized pure state $\ket{\phi}$ (with $\braket{\phi|\phi} =1$) 
towards that of the fully connected network, for all $p$, in the thermodynamic limit,
\begin{eqnarray}
   \lim_{N\to\infty} \bra{\phi} H_p(h)-H_{1}(h) \ket{\phi}
   &= \lim_{N\to\infty}\left[  \bra{\phi} \left( \sum_{i,j\in E_1} \frac{N}{|E_1|} \sigma^{z}_i\sigma^{z}_j - 
   \sum_{i,j\in E_{p}} \frac{N}{|E_{p}|}  \sigma^{z}_i\sigma^{z}_j \right) \ket{\phi} \right]  \nonumber\\
               &= \lim_{N\to\infty} \mathcal{O}(N^{-1/2}) = 0   \label{eq:diffconverge}
\end{eqnarray}
and its $\alpha$\textsuperscript{th} moment (\ref{app:convproof})
\begin{eqnarray}
   \lim_{N\to\infty} \bra{\phi} (H_p(h)-H_{1}(h))^\alpha \ket{\phi}
   &= \lim_{N\to\infty} \mathcal{O}(N^{-\alpha/2}) = 0 .   \label{eq:diffsqconverge}
\end{eqnarray}
Therefore, by letting $\ket{\phi}$ to be $\ket{\phi_p}$, the normalized eigenstate of $H_p(h)$ for finite $N$, 
the eigenstates of the model for any $h$ converge towards to that of the analytically tractable fully connected limit.
As a result, as discussed in \cite{tindallQuantumPhysicsConnected2022,searleThermodynamicLimitSpin2024}, 
the model undergoes the quantum phase transition at $\hec(p)=2$ for any value of $p$.

Shown in \figref{fig:eqPD} are the values of the order parameter $\mean{\Theta^{z}}$ 
for the model with $p=0.5$ (red triangles, computed with density matrix renormalization group algorithm (DMRG)
\cite{whiteDensityMatrixFormulation1992,
   schollwockDensitymatrixRenormalizationGroup2011,fishmanITensorSoftwareLibrary2022}), 
compared with the LMG model (green lines, ED) for various values of $N$ including 
$N=\infty$ (black dotted line, analytical) for different values of $h$.
Here $\mean{\cdot}$ denotes an averaging over network realizations of the quantum expectation values. 
As expected, the order parameter for $p=0.5$ (red triangles) shows convergence towards the exact ground state for $p=1$ 
(green lines) 
as increasing the network size, 
and shows excellent agreement with the $p=1$ for $N=100$.
However, even for $N=100$, there exists a notable discrepancy near the critical point for both $p=0.5$ and $p=1$.
This is due to the polynomial growth of the connectivity of a network, 
and as a result, the model exhibits a strong finite size effect in comparison to the finite-dimensional counterparts 
of the model (cf.~the results in \cite{suzuki2012quantum} for 1D and \cite{hashizumeDynamicalPhaseTransitions2022} for 2D). 
For $p=1$, this discrepancy persists even for $N=5000$, but the region shrinks as the network size increases.

\subsection{Proof of the convergence of the eigenstates of $H_p(h)$}
\label{app:convproof}
Let a state $\ket{\psi}$ be a normalized state ($\braket{\psi|\psi}$=1).
For any given $\ket{\psi}$, as proven in \cite{tindallQuantumPhysicsConnected2022}, we have
\begin{eqnarray}
   \bra{\psi} \left(H_p(h) - H_{1}(h)\right) \ket{\psi} &=
   -J\bra{\psi} \left(\sum_{i,j\in E_p} \frac{N}{|E_p|} \sigma^{z}_i\sigma^{z}_j - 
   \sum_{i,j\in E_{p=1}} \frac{N}{|E_{1}|}  \sigma^{z}_i\sigma^{z}_j\right) \ket{\psi} \nonumber\\
                                       &= \mathcal{O}(N^{-1/2}), \label{eqn:diffconverge}
\end{eqnarray}
where $H_p(h)$ is the Hamiltonian of TFIM-ER as defined in equation \eqnref{eq:model_ising} 
in the main text and $h$ is the transverse field strength.
Other symbols are as they are defined in the main text. 

Now let $\ket{\phi_p}$ be an arbitrary normalized eigenstate of Hamiltonian $H_p(h)$ with the eigenvalue $E_p$. 
With this, in this section, we show the following for arbitrary $0<p'\leq 1$
\begin{equation}
   \lim_{N\to\infty} \bra{\phi_p}\left(H_{p'}(h) - E_p\right)^{\alpha}\ket{\phi_p} = 
   \lim_{N\to\infty} \mathcal{O}\left( N^{-\alpha/2}  \right) = 0 \label{eqn:var_converge}
\end{equation}
where $\alpha$ is an integer greater than 1.

As $\alpha=1$ follows trivially from equation \eqnref{eqn:diffconverge}, we show that 
equation \eqnref{eqn:var_converge} holds for $1\leq \alpha$ by mathematical induction. 
First we show that convergence rate for $\alpha=2$ is at most $\mathcal{O}\left( N^{-2/2=-1}  \right)$ by 
letting $H_{p'}(h)=H_{p}(h) + (H_{p'}(h) - H_{p}(h))$, and define $\Delta_{\sigma^{z}\sigma^{z}}$ as 
\begin{equation}
   \Delta_{\sigma^{z}\sigma^{z}} 
   = -\sum_{i,j\in E_{p'}} \frac{JN}{|E_{p'}|} \sigma^{z}_i\sigma^{z}_j 
   + \sum_{i,j\in E_{p}} \frac{JN}{|E_{p}|}  \sigma^{z}_i\sigma^{z}_j,
\end{equation}
then 
\begin{equation}
   H_p' = H_p + \Delta_{\sigma^{z}\sigma^{z}}.
\end{equation}
Substituting this to the left-hand side of equation \eqnref{eqn:var_converge} gives
\begin{eqnarray}
   \lim_{N\to\infty} \bra{\phi_p}\left(H_{p'}(h) - E_p\right)^{2}\ket{\phi_p} &= 
   \lim_{N\to\infty} \left[ \bra{\phi_p}\Delta_{\sigma^{z}\sigma^{z}}^2\ket{\phi_p}
   \ket{\phi_p}^2 \right] \label{eqn:var_converge2}
\end{eqnarray}
This expectation value also converges like $\mathcal{O}\left( N^{-1}  \right)$.  
Let $\ket{\phi}$ be a superposition of basis states in $z$-axis, $\ket{{\bm \sigma_l}}$, 
with complex parameter $c_l$
\begin{equation}
   \ket{\phi_p}=\sum_l c_l\ket{{\bm\sigma}_l}. 
\end{equation}
By inserting the identity, we obtain the correct limit 
\begin{eqnarray}
   \left| \bra{\phi_p} \Delta_{\sigma^{z}\sigma^{z}}^2 \ket{\phi_p} \right|
   & = \left| \sum_l c_l^* \bra{{\bm\sigma}_l}\Delta_{\sigma^{z}\sigma^{z}}
   \sum_m \ket{{\bm \sigma}_m}\bra{{\bm\sigma}_m}\Delta_{\sigma^{z}\sigma^{z}} 
   \sum_n c_n\ket{{\bm\sigma}_n}
   \right|
   \nonumber\\
   &= \mathcal{O}(N^{-1}) \sum_l c_l^*c_l  = \mathcal{O}(N^{-1}),
\end{eqnarray}
and hence the $\mathcal{O}(N^{-1})$ convergence of equation \eqnref{eqn:var_converge2} and hence 
equation \eqnref{eqn:var_converge} for $\alpha=2$ 
is proven.

Now, let us assume that equation \eqnref{eqn:var_converge} converges like $\mathcal{O}\left( N^{\alpha/2}  \right)$ for $\alpha=k$. 
Then for $\alpha=k+1$, we have
\begin{eqnarray}
   &\lim_{N\to\infty} \bra{\phi_p}\left(H_{p'}(h) - E_p\right)^{k+1}\ket{\phi_p} \nonumber\\
   &=\lim_{N\to\infty} \bra{\phi_p}\left(H_{p'}(h) - E_p\right) \left(H_{p'}(h) - E_p\right)^{k}\ket{\phi_p} 
\end{eqnarray}
We then substitute an identity $\mathbb{I}=\sum_q \ket{\phi_{p,q}}\bra{\phi_{p,q}}$
where $\ket{\phi_{p,q}}$ the $q$\textsuperscript{th} eigenstate of the Hamiltonian $p$, 
and we define $\ket{\phi_p}=\ket{\phi_{p,0}}$
\begin{eqnarray}
   &\lim_{N\to\infty} \bra{\phi_p}\left(H_{p'}(h) - E_p\right) \left(H_{p'}(h) - E_p\right)^{k}\ket{\phi_p}  \nonumber\\
   &=\lim_{N\to\infty} \bra{\phi_p}\left(H_{p'}(h) - E_p\right) \mathbb{I} \left(H_{p'}(h) - E_p\right)^{k}\ket{\phi_p}.
   \label{eqn:withidentity}
\end{eqnarray}

We evaluate $\bra{\phi_{p,0}} \Delta_{\sigma^{z}\sigma^{z}}\ket{\phi_{p,q}}=C\delta_{0,q}$ where $C$ 
is $\mathcal{O}\left( N^{-1/2}  \right)$ parameter and $\delta_{q,r}$ is a Kronecker delta. 
Let 
\begin{equation}
   \ket{\phi_{p,q}}=\sum_l c_{l,q}\ket{{\bm\sigma}_l}, 
\end{equation}
then 
\begin{eqnarray}
   &\bra{\phi_{p,0}} \Delta_{\sigma^{z}\sigma^{z}}\ket{\phi_{p,q}} \nonumber\\
   &= 
   \sum_{i,j,l,m} c_{l,0}\bra{{\bm\sigma}_l}  
   \left(-\sum_{i,j\in E_{p'}} \frac{JN}{|E_{p'}|} \sigma^{z}_i\sigma^{z}_j 
   + \sum_{i,j\in E_{p}} \frac{JN}{|E_{p}|}  \sigma^{z}_i\sigma^{z}_j\right)
   c_{m,0}\ket{{\bm\sigma}_l} \nonumber\\
   &= 
   -\sum_{i,j,l,m} \left(\sum_{i,j\in E_{p'}} \frac{JN}{|E_{p'}|} -\sum_{i,j\in E_{p}} \frac{JN}{|E_{p}|}\right)
   \left(c_{l,0}\bra{{\bm\sigma}_l} \right)
   \left(c_{m,0}\ket{{\bm\sigma}_l} \right) \nonumber\\
   &= -\delta_{0,q} \sum_{i,j} \left(\sum_{i,j\in E_{p'}} \frac{JN}{|E_{p'}|} -\sum_{i,j\in E_{p}} \frac{JN}{|E_{p}|}\right)
   = C\delta_{0,q}. 
\end{eqnarray}
With this result, equation \eqnref{eqn:withidentity} evaluates to 
\begin{eqnarray}
   &\lim_{N\to\infty} \bra{\phi_p}\left(H_{p'}(h) - E_p\right) \mathbb{I} \left(H_{p'}(h) - E_p\right)^{k}\ket{\phi_p}
   \nonumber\\
   &=
   \lim_{N\to\infty} C \ket{\phi_{p}}\bra{\phi_p} 
   \left(H_{p'}(h) - E_p\right)^{k}\ket{\phi_p} = \lim_{N\to\infty} C \mathcal{O}\left(N^{-k/2}\right) \nonumber\\
   &= 
   \mathcal{O}\left(N^{-(k+1)/2}\right)
\end{eqnarray}
As expected. 
Therefore, the eigenstates of the model converges to their fully connected counterpart in the thermodynamic limit, 
and the deviations from the fully connected limit of higher order correlations vanish much faster than the deviation of the mean. 

\section{Time averaged order parameter in the thermodynamic limit}
\label{app:timeaverage}
In this appendix, we derive the time-averaged value of the order parameter in the thermodynamic limit ($N\to\infty$). 
We start from the scaled mean-field Hamiltonian of the fully connected limit given by equation \eqnref{eq:p1H} 
\begin{equation}
   H = H_1(h)/N = -J \braket{\Theta^{z}}^2 - h\sqrt{1-\braket{\Theta^{z}}^2}\cos(2k),
\end{equation}
where $\braket{\Theta^{\alpha}}=\sum_i\braket{\sigma^{\alpha}_i}/N$ for $\alpha\in(x,y,z)$ and 
$k=\arctan(\braket{\Theta^{y}}/\braket{\Theta^{x}})/2$. 
From the conservation of energy, for $h_f=1$ and $\braket{\Theta^{z}}=1$ at $t=0$, 
$\braket{\Theta^{z}(t)}$ and $k(t)$ has a following relation
\begin{equation}
   \braket{\Theta^z(t)} = \sqrt{1-(h/J)^2\cos^2(2k(t))}
\end{equation}

We first calculate the period $T$ of an oscillation. 
It is is twice the time it takes for $k$ to go from $\pi/4$ to $0$. 
Therefore, we obtain the period 
\begin{eqnarray}
   T \label{eqn:theo_period}
   &=2\int_{0}^{T/2}dt = 2\int_{\pi/4}^{0} \frac{1}{\partial_k H} \deriv{\braket{\Theta^z}}{k}dk\\ 
   &=2\int_{\pi/4}^{0} \frac{1}{\sqrt{1-(h/J)^2\cos^2(2k)}}dk = K((h/J)^2)
\end{eqnarray}
where $K(m)$ is the elliptic integral of the first kind.

Similarly, we compute the total $\braket{\Theta^z}$ over the period
\begin{eqnarray}
   \Sigma^{z} = 2\int_{\pi/4}^{0} \braket{\Theta^z} \deriv{t}{k} dk
   = 2\int_{0}^{\pi/4} \frac{\braket{\Theta^z}}{\partial_{\braket{\Theta^z}} H} dk
   =\pi/2
\end{eqnarray}
Hence, we obtain the time-averaged order parameter, $\overline{\overline{\braket{\Theta}^{z}}}$, in the thermodynamic limit, 
\begin{equation}
   \overline{\overline{\braket{\Theta}^{z}}}  = \frac{\pi}{2K((h/J)^2)}
\end{equation}

\newpage
\section{Proof of the convergence of the wave function evolution}
\label{app:convdynamicsproof}
In this appendix, we show that evolution of the wave functions becomes identical for the different $p$
evolved from the common initial state $\ket{\psi(0)}$. 
We show this by how the fidelity $\mathcal{F}=\braket{\psi_p(t)|\psi_1(t)}$ evolves for finite $N$,
where $\ket{\psi_p(t)}=\exp(-iH_pt)\ket{\psi(0)}$, $\ket{\psi_1(t)}=\exp( -iH_1t )\ket{\psi(0)}$, 
and $H_p$ is a Hamiltonian of TFIM on \erdosrenyi{} network as defined in the main text. 

The time derivative of $F$ is
\begin{equation}
   \frac{\partial}{\partial t} \mathcal{F} = i\bra{\psi_p(t)}\left(H_p - H_1\right)\ket{\psi_1(t)}
\end{equation}
We now expand each wave function in terms of the basis state in $z$-axis, $\ket{\bm{\sigma}}$ 
\begin{equation}
   \ket{\psi_p(t)} = \sum_{m} C_{m}(t,p)\ket{\bm{\sigma}}, 
\end{equation}
where $C(t,p)=\braket{\bm{\sigma}|\psi_p(t)}$ are complex coefficients. 
Like in \ref{app:convproof} we define
\begin{equation}
   \Delta_{\sigma^z\sigma^z} = 
   -\sum_{i,j\in E_1} \frac{JN}{|E_{p'}|} \sigma^{z}_i\sigma^{z}_j 
   + \sum_{i,j\in E_{p}} \frac{JN}{|E_{p}|}  \sigma^{z}_i\sigma^{z}_j
\end{equation}
then, 
\begin{equation}
   \bra{\psi_p(t)} \Delta_{\sigma^z\sigma^z} \ket{\psi_{1}(t)}
   = \sum_{m,n} C^{*}_{m}(t,p) C_{n}(t,1) 
   \bra{\bm{\sigma_{m}}} \Delta_{\sigma^z\sigma^z} \ket{\bm{\sigma_{n}}}
   \delta_{m,n}
\end{equation}
where $\delta_{m,n}$ is Kronecker's delta.
The magnitude of the overlap, therefore, can be bounded from the above 
\begin{eqnarray}
   0 &< \bigg| \deriv{}{t}\mathcal{F} \bigg|=\bigg| \bra{\psi_p(t)} \Delta_{\sigma^z\sigma^z}\ket{\psi_1(t)} \bigg|  \nonumber\\
   &< \bigg| \sum_{m} C^{*}_{m}(t,p) C_{m}(t,1) \bigg|  
   \sup \left\{\bigg| \bra{\bm{\sigma^{\alpha}}} \Delta_{\sigma^z\sigma^z} \ket{\bm{\sigma^{\alpha}}} \bigg|\right\} \nonumber\\
   &=  \mathcal{O}\big(N^{-1/2}\big). 
\end{eqnarray}
Hence, in the thermodynamic limit, the model possesses the same quench dynamics for all the values of $p$ ($0<p\leq 1$).

\section{Mean field equations of motion}
\label{app:mfem}
To perform the semiclassical and mean-field simulations, we first derive the mean-field equations of motion. 
Starting from the Ehrenfest equations of the quantum mechanical observables 
\begin{equation}
   i\deriv{}{t}\braket{\sigma^{\alpha}_j} =  \braket{\left[\sigma^{\alpha}_j, H_p(h) \right]}. 
\end{equation}
we apply the mean-field approximation
$\braket{\sigma^{\xi}_i\sigma^{\zeta}_j}\approx \braket{\sigma^{\xi}_i}\braket{\sigma^{\zeta}_j}$
and obtain the mean-field equations of motion 
\begin{equation}
   \begin{array}{ll}
      \frac{d}{dt}\expval{\sigma_i^x} &= \frac{2NJ}{|E_p|} \sum_{j=1}^N A_{ij}(\ergraph(N,p)) \bigexpval{\sigma_i^y} \bigexpval{\sigma_j^z}\;, \\ [\jot]
      \frac{d}{dt}\expval{\sigma_i^y} &= -\frac{2NJ}{|E_p|} \sum_{j=1}^N A_{ij}(\ergraph(N,p)) \bigexpval{\sigma_i^x} \bigexpval{\sigma_j^z} 
      +2 h \bigexpval{\sigma_i^z} \;, \\ [\jot]
   \frac{d}{dt}\expval{\sigma_i^z} &= -2 h \bigexpval{\sigma_i^y}\;, \label{eq:SCeqmotion}
   \end{array}
\end{equation}
where $A_{ij}(G)$ is the adjacency matrix of a network $G$ \cite{huangSymmetrybreakingDynamicsFinitesize2018}. 

\section{System size dependence of the rate function}
\label{app:lambda_vs_N}
\begin{figure}
   \includegraphics[width=1.0\textwidth]{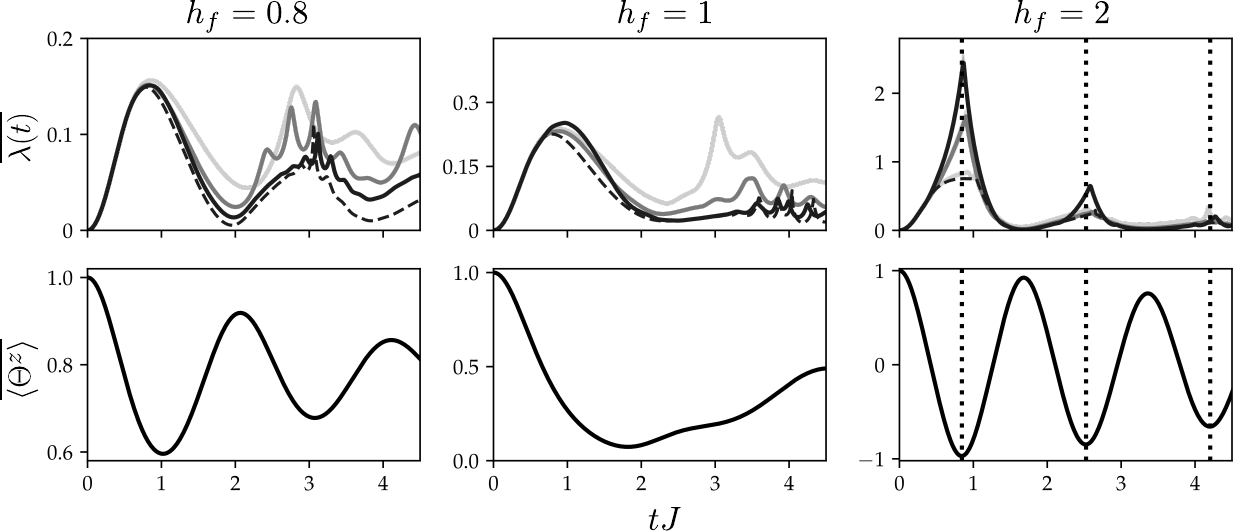}
   \caption{{\bf Rate function for different system sizes.}
      Rate function for system sizes $N=20$, $50$, and $100$ (light to dark) for $p=0.5$ for quenches 
      $h_f=0.6$ (left), $h_f=1$ (middle), and $h_f=2$ (right).
            Plotted in the bottom panels are $\mean{\Theta^{z}}$ for $p=0.5$; 
      for $h_f=2$, the turning points of $\mean{\Theta^{z}}$ are indicated with vertical dotted lines. 
      The simulations are conducted with TDVP algorithm with MPS with the maximum bond dimension $\chi=200$ and $\Delta t=0.01$. 
      The rate function is computed from the numerically obtained Loschmidt amplitude $\mathcal{G}(t)$.
      The error bars are plotted as shaded regions, but they are too small to be visible. 
      For $p=1$, $\lambda(t)$ is calculated exactly to 200 significant figures \cite{MATLAB}.\label{fig:fsrf}}
\end{figure}
In this appendix we show the system size dependence of the rate function for the quenches $h_f=0.6$, $1$ and $2$
explored in \secref{subsec:DQPT-II}.
Shown in \figref{fig:fsrf} is the rate function for system sizes $N=20$, $N=50$, and $100$ for $p=0.5$ for quenches 
$h_f=0.6$ (left), $h_f=1$ (middle), and $h_f=2$ (right).
For a quench below the critical point, rate function converges towards the $p=1$ limit with the system size.
However, for the quench above the critical point ($h_f=2$, right), the rate function diverges from the limit, and show 
strong divergence near the turning point of the order parameter. 
For $h_f=1$, it admits small deviation from the limit while the rate function converges towards the limit in the later times. 

\section{Influence of the fluctuations to the higher-order moments}
\label{app:flucandcumul}

\begin{figure}[tb]
   \includegraphics[scale=1.0]{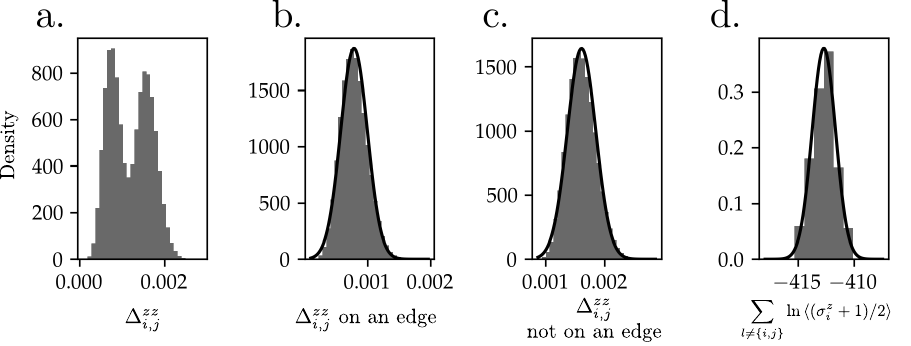}
   \caption{
      {\bf Statistics of product of local operators and joint $2$\textsuperscript{nd} moment 
      for $\bm{\ergraph(100,0.5)}$ at $ \bm{t=0.84}$ for quenches with $\bm{h_f=2}$.}
      {\bf a.}~A numerically obtained distribution of 
         $\Delta^{zz}_{i,j} = \braket{\sigma^{z}_i\sigma{^z}_j} - \braket{\sigma^{z}_i}\braket{\sigma^{z}_j}$.
      {\bf b.}~(a) for the sites where there is an edge between $i$ and $j$. 
      {\bf c.}~(c) for the sites where there is no edge between $i$ and $j$.
      {\bf d.}~A numerically obtained distribution of $\sum_{l\neq \{i,j\}} \ln \braket{(\sigma^{z}_l + 1)/2}$,
      where the summation goes over the vertices that are not involved in the join $2$\textsuperscript{th}-order moment.
      Observed normal distribution implies that a product $\prod_{l\neq \{i,j\}} \braket{(\sigma^{z}_l + 1)/2}$
      follows the log-normal distribution \cite{yangNormalLognormalMixture2008}.
   } \label{fig:statfig}
\end{figure}

In this appendix we show that if the covariance between joint $m$\textsuperscript{th}-order moment
and the product $\prod_{l\neq\{i,j,k\cdots\}} \braket{\frac{\sigma_l+1}{2}}$ in equation \eqnref{eqn:rfapprox} of the main text
does not vanish faster than $N^{-m}$, then it influences the behaviour of the average rate function, $\overline{\lambda(t)}$. 
The central limit theorem 
tells us that $i,j,k\cdots$ go over the all combinations of $m$ non-overlapping vertices $i\neq j\neq k \neq\cdots$. 
We assume $\Delta_{i,j,k \cdots}^{zzz\cdots}$ 
with different indices are drawn independently from a normal distribution with mean $\mu_m(\mathcal{C})$ 
and variance $\sigma^2_{m}(\mathcal{C})$ 
after resolving them over the possible edge configurations $\mathcal{C}$ (\figref{fig:statfig}a--c). 
The distribution of a sum of random variables $\sum_l \ln((\sigma_{l}+1)/2)$, on the other hand, 
follows a normal distribution with mean $(N-m)\mu_0$ and variance $(N-m)^2\sigma_0^2$ due to the central limit theorem,
where $\mu_0$ and $\sigma_0$ are the mean and the standard deviation of the distribution of $\braket{\ln (\sigma^{z}_i+1)/2}$ 
at different sites $i$ over the whole ensemble.
Hence the distribution of $\prod_{l \neq i,j,k,\cdots} \braket{\frac{\sigma_l^{z}+1}{2}}$ on an ensemble 
follows a log-normal distribution (\figref{fig:statfig}d).

The mean of a product of a normal and log-normal distribution is sensitive to the correlation between 
the two distributions \cite{yangNormalLognormalMixture2008}. 
Thus, $\mean{|\mathcal{G}(t)|^2}$ is approximated as
\begin{eqnarray}
   \mean{|\mathcal{G}(t)|^2} &\approx 
   \mathbb{E} \left[ \prod_{l} \braket{\frac{\sigma_l^{z}+1}{2}} \right] + 
   \sum_m 
   \frac{~_NC_m}{{2^m}}
   \mathbb{E} \left[
      \Delta_{i,j,k\cdots}^{zzz\cdots} \prod_{l \neq i,j,k,\cdots} \braket{\frac{\sigma_l^{z}+1}{2}}
   \right]
   \nonumber\\
   &=
   e^{N\mu_0+\frac{N\sigma_0^2}{2}}
   \left(1+\sum_{m=2}\frac{~_NC_m }{2^m} \overline{C}_{m,\mathrm{approx}}\right)
   \label{eqn:corrtoexp}
\end{eqnarray}
where
\begin{eqnarray}
   \overline{C}_{m,\mathrm{approx}}
   &=
   \frac{e^{(N-m)\mu_0+\frac{(N-m)\sigma_0^2}{2}}}{e^{N\mu_0+\frac{N\sigma_0^2}{2}}}
      \sum_{\mathcal{C}} p(\mathcal{C})\mu_m(\mathcal{C}) +
 \sum_{\mathcal{C}} p(\mathcal{C})\sigma_{0,m}(\mathcal{C}) \nonumber\\
   &\approx \frac{\mu_m + \sigma_{0,m}}{\mean{(1+\sigma^{z}_i)/2}^m}, 
   \label{eqn:defCmapprox}
\end{eqnarray}
$\sigma_{0,m}(\mathcal{C})$ is the correlation between the distributions,
$p(\mathcal{C})$  is the probability of obtaining a configuration $\mathcal{C}$, 
$\mu_m=\sum_{\mathcal{C}} p(\mathcal{C})\mu_m(\mathcal{C})$ is the configuration resolved mean value of 
$\Delta^{zzz\ldots}_{i,j,k,\ldots}$, and 
$\sigma_{0,m}=\sum_{\mathcal{C}} p(\mathcal{C})\sigma_{0,m}(\mathcal{C})$ is the configuration resolved correlation.
Thus we obtain the approximated deviation $\Delta_m$ in \eqnref{eqn:Deltam} in the main text. 

Finally, using the above result, we argue that in the thermodynamic limit, 
\begin{equation}
\mean{\lambda(t)}=-\frac{1}{N}\ln \mean{|\mathcal{G}|^2}=\ln \mean{|\mathcal{G}|^2}^{-1/N}
\end{equation}
diverges to infinity when $\mean{(\sigma_i+1)/2}=0$. 
From \eqnref{eqn:corrtoexp}, we see that $\lim_{N\to\infty} \mean{|\mathcal{G}|^2}^{1/N}=0$ if 
\begin{equation}
   \left(1+\sum_{m=2}\frac{~_NC_m }{2^m} \overline{C}_{m,\mathrm{approx}}\right)^{1/N} \approx 
   \frac{\mean{|\mathcal{G}|^2}^{1/N}}
   {\mean{(\Theta + 1)/2}} \label{eq:gontp1}
\end{equation}
does not grow faster than $e^{-\mu_0}$,
where $e^{\mu_0} \approx \mean{(\Theta+1)/2}$ approaches the mean-field value of $0$ like $\mathcal{O}(N^{-1})$.
As shown \figref{fig:divergingcorrelation}b in the main text, it does not grow faster than $N^{1}$,
near the first turning point of the order parameter. 

\end{document}